\documentclass{article}

\usepackage{graphicx}
\usepackage{booktabs} 
\usepackage{tabularx} 
\usepackage{amsmath}  
\usepackage{caption}  
\usepackage{subcaption} 
\usepackage{geometry} 
\usepackage{float}

\usepackage{apacite}  
\usepackage{times}    
\usepackage{setspace} 

\usepackage{listings} 
\usepackage{xcolor}

\usepackage{listings} 
\usepackage{xcolor} 
\usepackage{geometry} 
\definecolor{codegreen}{rgb}{0,0.6,0}
\definecolor{codegray}{rgb}{0.5,0.5,0.5}
\definecolor{codepurple}{rgb}{0.58,0,0.82}
\definecolor{backcolour}{rgb}{0.95,0.95,0.92}

\lstdefinestyle{mystyle}{
    backgroundcolor=\color{backcolour},   
    commentstyle=\color{codegreen},
    keywordstyle=\color{magenta},
    numberstyle=\tiny\color{codegray},
    stringstyle=\color{codepurple},
    basicstyle=\ttfamily\footnotesize,
    breakatwhitespace=false,         
    breaklines=true,                 
    captionpos=b,                    
    keepspaces=true,                 
    numbers=left,                    
    numbersep=5pt,                  
    showspaces=false,                
    showstringspaces=false,
    showtabs=false,                  
    tabsize=2
}

\definecolor{codegreen}{rgb}{0,0.6,0}
\definecolor{codegray}{rgb}{0.5,0.5,0.5}
\definecolor{codepurple}{rgb}{0.58,0,0.82}
\definecolor{backcolour}{rgb}{0.95,0.95,0.92}

\begin{document}

\title{Practical programming research of Linear DML model based on the simplest Python code: From the standpoint of novice researchers}
\author{SHUNXIN YAO\\ QUT}
\date{}
\maketitle

\begin{abstract}
This paper presents linear DML models for causal inference using the simplest Python code on a Jupyter notebook based on an Anaconda platform and compares the performance of different DML models. The results show that current Library API technology is not yet sufficient to enable novice Python users to build qualified and high-quality DML models with the simplest coding approach. Novice users attempting to perform DML causal inference using Python still have to improve their mathematical and computer knowledge to adapt to more flexible DML programming. Additionally, the issue of mismatched outcome variable dimensions is also widespread when building linear DML models in Jupyter notebook.
\end{abstract}

\section{Introduction}
Causal inference is crucial in data science because it reveals causal relationships between variables, not just correlations. This is essential for decision-making, policy evaluation, and scientific research. Through causal inference, researchers can assess the effectiveness of interventions, providing a scientific basis for resource allocation and strategic planning (Pearl, Glymour, \& Jewell, 2016). However, traditional statistical methods have limitations in causal inference, such as inability to determine causality direction, reliance on hypothesis testing, and difficulty controlling confounding variables, which can lead to misleading conclusions. To address these challenges, double machine learning (DML) has emerged. DML combines machine learning with causal inference, effectively handling the estimation of causal effects in high-dimensional data. Its core steps include feature modeling, causal effect estimation, and double validation, using machine learning models to control confounding variables, reduce estimation bias, and enhance the robustness of results. The flexibility and adaptability of DML enable it to handle complex nonlinear relationships, expanding the scope of causal inference applications and bringing new opportunities and challenges to data science (Chernozhukov et al., 2018).

As Python, a powerful open-source "glue code", becomes more widely adopted, data analysts are increasingly using Python as their primary tool for data analysis. Python programming courses based on Jupyter notebooks of Anaconda platform have become popular in modern colleges and universities around the world. However, in the linear DML domain, which serves statistical inference, applying scientific DML theory to practical data analysis programming still presents significant challenges. This article, from the perspective of a beginner in Python data analysis, attempts to use simplified Python code and existing library APIs to establish and compare the performance of popular machine learning models within the DML framework. It is important to note that this performance is based on model building results at the beginner level and does not represent general data analysis proficiency.

\section{Literature Review}
Propensity Score Matching (PSM) and Instrumental Variable (IV) were two commonly used methods in causal inference before. PSM matches treatment and control groups by estimating the probability of individuals being treated, reducing selection bias. However, it relies on the accuracy of the model, cannot control unobservable confounding factors, and may degrade matching quality in high-dimensional data. The IV method addresses endogeneity by introducing instrumental variables that are correlated with the treatment variable but only with the outcome variable through the treatment variable. However, selecting instrumental variables and meeting exogeneity assumptions can be challenging, and weak instrument problems may affect the validity of estimates. Therefore, despite their significant applications in controlling confounding and endogeneity, researchers must carefully consider the limitations of PSM and IV. To address these shortcomings of traditional methods, DML models have emerged (Fuhr, Berens, \& Papies, 2024).

Chernozhukov et al. (2018) proposed the Double Machine Learning (DML) method, which is used to estimate treatment effects and structural parameters in causal inference for high-dimensional data. By combining machine learning with semi-parametric methods, DML can eliminate model bias, providing more accurate and unbiased estimates. They also demonstrated the application of this method in econometrics, particularly in addressing parameter estimation issues in policy evaluation and economic models.

Bach et al. (2022) introduced a Python open-source library for double machine learning (DML), designed to achieve robust causal inference through Neyman orthogonality, machine learning methods, and sample partitioning. It adopts object-oriented programming (OOP), supports various causal models (such as partially linear regression and instrumental variable regression), and provides a flexible API for extensible functionality. Compared to EconML and CausalML, DML places greater emphasis on the validity of orthogonality conditions and statistical inference.

Chernozhukov et al. (2024) further proposed the Double Machine Learning (DML) method, which combines the orthogonality of Neyman and cross-fitting to reduce the impact of high-dimensional interference parameter estimation errors on causal inference, achieving $\sqrt{N}$ convergence in target parameter estimation while maintaining statistical inference validity. DML is applicable to partial linear regression, instrumental variable models, and treatment effect estimation, providing a robust methodological foundation for high-dimensional causal inference.

Based on the theory and practice of modern DML, Flores \& Chernozhukov (2018) established an open-source code library for the GitHub project, demonstrating how to implement DML in Python for causal inference and effect estimation. It provides detailed documentation and examples, integrating machine learning methods such as random forests and Lasso regression, and supports model evaluation and result visualization, making it suitable for causal inference analysis of high-dimensional data.

\section{Theoretical Framework}
\subsection{Theory Overview}
Double Machine Learning (DML) is a method used to estimate causal effects, particularly suitable for situations with high-dimensional confounding variables. The core idea of DML is to use machine learning models to separately estimate the conditional expectations of the treatment variable and the outcome variable, thereby eliminating confounding bias.

The basic principle of DML can be divided into the following steps:

\begin{table}[H]
\centering
\caption{Steps in Double Machine Learning}
\begin{tabularx}{\textwidth}{lX} 
\toprule
Step 1. & Conditional expectation estimation. First, use the machine learning model to estimate the conditional expectations of the treatment variable $T$ and the outcome variable $Y$ respectively. Specifically, estimate $E[T|X]$ and $E[Y|X]$, where $X$ is the confounding variable. \\
\midrule
Step 2. & Residual calculation. The effect of confounding variables is removed by calculating the residual of the treatment variable and the outcome variable. The residual of the treatment variable is $T - E[T|X]$, and the residual of the outcome variable is $Y - E[Y|X]$. \\
\midrule
Step 3. & Causal effect estimation. The causal effect is estimated using the residual. By regressing the relationship between the residuals, the causal effect of the treatment variable on the outcome variable can be obtained. \\
\bottomrule
\end{tabularx}
\end{table}

\subsection{Mathematical Principles}
The core formula of DML can be simplified as follows:

\begin{equation}
\tilde{Y} = \theta \tilde{T} + \epsilon
\end{equation}

where:

\begin{itemize}
\item $\tilde{Y} = Y - \hat{E}[Y \mid X]$ is the residual of the result variable.
\item $\tilde{T} = T - \hat{E}[T \mid X]$ is the residual of the intervention variable.
\item $\theta$ is the estimated value of the causal effect.
\item $\epsilon$ is the error term.
\end{itemize}

The principle of this formula is to remove the influence of confounding variables $X$, making the relationship between the intervention variable $T$ and the outcome variable $Y$ purer. By analyzing the relationship between regression residuals, it can more accurately estimate the causal effect $\theta$ of the intervention variable on the outcome variable (Chernozhukov et al., 2018). In summary, DML uses a double machine learning model to separately estimate the conditional expectations of the intervention variable and the outcome variable, then eliminates confounding biases through residual calculation and regression analysis, ultimately obtaining an estimate of the causal effect.

\section{Data}
\subsection{Data Sources}
The dataset is random data generated by the Python NumPy library, with no external source of actual data. A random number generator was used to create variables $X$ and response variable $y$. The data is entirely simulated to create a hypothetical regression model.

\subsection{Data Characteristics}
The dataset includes five features including three independent variables ($X_1$ to $X_3$) and two control variables ($X_4$ to $X_5$) and one dependent variable ($y$). The mean of the bunch of independent variables and control variables is close to 0, with standard deviations ranging from 0.96 to 1.03, indicating that the distribution of the independent variables and control variables has some dispersion and significant data fluctuation, with minimum and maximum values ranging from -3.17 to 3.93. The distribution of these independent variables and control variables generally conforms to the characteristics of a standard normal distribution, and the quartiles show a relatively even distribution of data.

The mean of the dependent variable $y$ is 0.0904, with a standard deviation of 4.1033, indicating significant variability. The minimum and maximum values of $y$ are -14.36 and 13.44, respectively, showing a wide range of fluctuations. The quartile results indicate that the distribution of $y$ is relatively scattered, suggesting that changes in the data may be due to noise or other factors. These statistical characteristics help understand the distribution of the data and provide useful information for subsequent analysis.

\begin{figure}[H]
\centering
\includegraphics[width=0.8\textwidth]{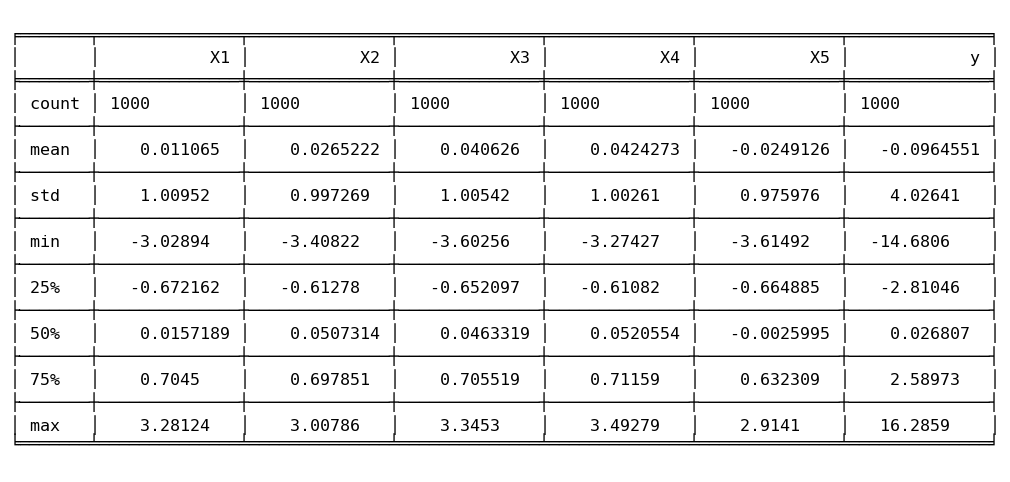}
\caption{Descriptive Statistics of Data} 
\end{figure}

\begin{figure}[H]
\centering
\includegraphics[width=0.8\textwidth]{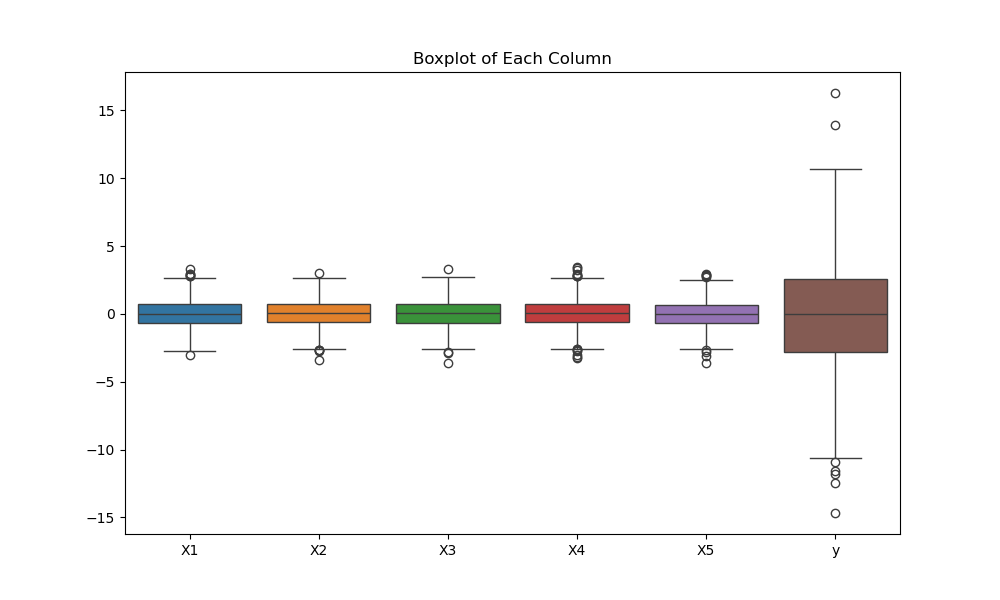}
\caption{Boxplot of Data}
\end{figure}

\begin{figure}[H]
\centering
\includegraphics[width=0.8\textwidth]{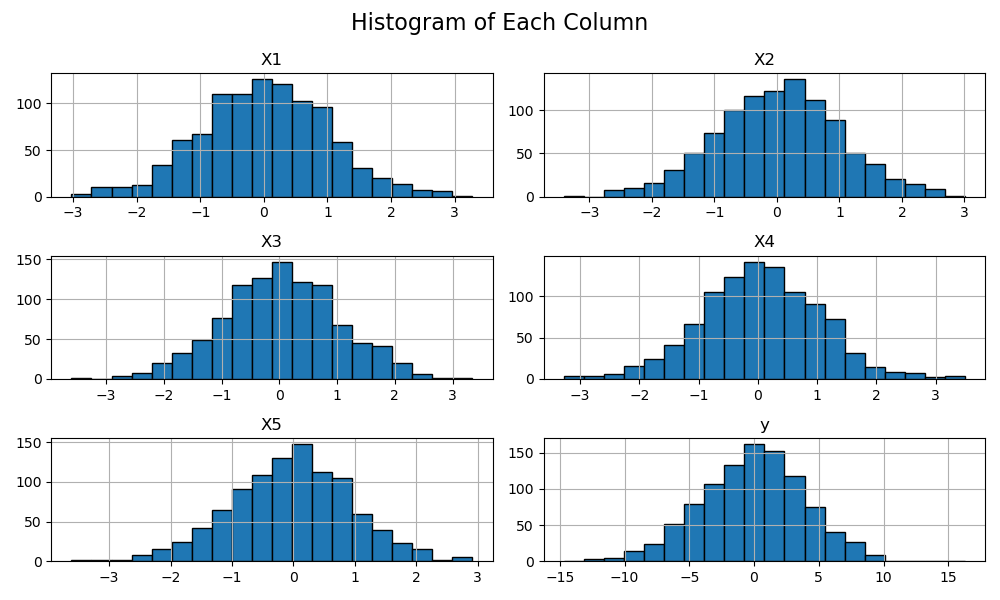}
\caption{Histogram of Data} 
\end{figure}

\section{Methodological Description}
\subsection{Model Selection}
The model selected in this paper can be implemented on an Anaconda Jupyter notebook using the simplest Python code. These models also hold extreme theoretical importance. The remaining important models were not listed due to a data dimension bug in the outcome variable, causing other significant machine learning models to fail during runtime. This bug has a common negative impact under current data packages, making how to address this issue a crucial challenge for the future.

This study did not use the ready-made GitHub database but used the simplest code to call the ordinary library API for research to evaluate the practical programming situations of novice researchers.

\subsection{Model Establishment}
Variable Definition:
\begin{table}[H]
\centering
\caption{Variable Definitions}
\begin{tabularx}{\textwidth}{lX} 
\toprule
Variable & Variable Description \\
\midrule
Independent variable ($T$) & An array of size (1000,5) is generated through \texttt{np.random.randn(1000,5)} to represent five features of 1000 samples. The first three are independent variables. $X$ and $T$ together form a matrix ($X_1$ to $X_5$) containing 1000 samples and 5 features. Each feature is a random number drawn from a standard normal distribution, representing five different variables. \\
\midrule
Covariance ($X$) & An array of size (1000,5) is generated through \texttt{np.random.randn(1000,5)}, which represents 5 features of 1000 samples. The last two are covariances. $X$ and $T$ together form a matrix ($X_1$ to $X_5$) containing 1,000 samples and five features. Each feature is a random number drawn from a standard normal distribution, representing five different variables. \\
\midrule
Dependent variable ($y$) & $y$ is the response variable (i.e., the target variable), which is obtained through a linear combination of the independent variable $X$ and a given set of coefficients [1.5, -2, 0.5, 0, 3]. A noise term \texttt{np.random.randn(1000)} is added to simulate real-world data errors, making the model imperfect and thus more closely resembling real-world data. \\
\bottomrule
\end{tabularx}
\end{table}

The basic formula is:

\begin{equation}
y = X \cdot \beta + \epsilon
\end{equation}

where $\beta = [1.5, -2, 0.5, 0, 3]$ is the coefficient of the linear regression model, and $\epsilon$ is the noise from the standard normal distribution.

This study uses the Double Machine Learning (DML) method to estimate causal effects. The specific steps are as follows:

\begin{table}[H]
\centering
\caption{Steps in DML Model Implementation}
\begin{tabularx}{\textwidth}{lX} 
\toprule
Step 1. & Data generation and preprocessing. I first generated a synthetic dataset containing 1,000 samples and 5 features using NumPy. The response variable $y$ was generated by linearly combining each column of the feature matrix $X$ with corresponding weights and adding noise. Subsequently, the dataset was divided into a training set (80\%) and a test set (20\%), and processing variables $T$ and control variables $X$ were selected based on the features. \\
\midrule
Step 2. & Model initialization. Taking the random forest as an example, I implemented the DML model using the \texttt{LinearDML} from the \texttt{econml} library and selected the random forest regressor (RF) as the prediction model. Specifically, RF is used for predicting the outcome variable $y$ and the treatment variable $T$. To better control the variables, I chose the nonlinear first-stage model in my model setup (i.e., \texttt{linear\_first\_stages=False}). Other machine learning models were tried in turn, and the results of the models that ran smoothly were reported. \\
\midrule
Step 3. & Model training and effect evaluation. Model training fits the relationship between outcome variable $y$ and treatment variable $T$, while control variable $X$ is used to adjust for confounding effects. After training, the test set is predicted using the \texttt{effect} method to obtain estimates of causal effects. To evaluate model performance, I calculated common regression metrics such as mean squared error (MSE), mean absolute error (MAE), and $R^2$. Additionally, the time required for the training process was recorded. \\
\midrule
Step 4. & Result output. Through the evaluation of the model, the prediction effect index of the model is obtained, including MSE, MAE and $R^2$, and the time consumption of the model training is presented. \\
\bottomrule
\end{tabularx}
\end{table}

\subsection{Model Results}
\begin{table}[H]
\centering
\caption{Model Performance Comparison}
\begin{tabular}{lllll}
\toprule
 & Training time (s) & MSE & MAE & $R^2$ \\
Random Forest & 0.3418 & 14.4012 & 2.9868 & -0.0131 \\
MLP & 6.1360 & 16.9532 & 3.3011 & -0.0026 \\
XGBoost & 1.7447 & 14.6655 & 3.0486 & -0.0146 \\
CatBoost & 0.7989 & 17.2292 & 3.3485 & -0.0065 \\
Lasso & 0.1200 & 15.9975 & 3.2034 & -0.0050 \\
Ridge & 0.0084 & 14.1610 & 2.9575 & 0.0038 \\
OLS & 0.0072 & 0.9749 & 0.7898 & 0.9314 \\
SVM & 0.0441 & 1.6339 & 0.9180 & 0.8851 \\
\bottomrule
\end{tabular}
\end{table}

According to the data in the table, there are significant differences in performance across various regression models on different metrics. First, from the perspective of training time, OLS, Ridge, and Lasso have shorter training times, making them suitable for use when computational resources are limited. Among these, OLS has the shortest training time at just 0.0072 seconds, while MLP has the longest training time at 6.1360 seconds.

In terms of model prediction accuracy, the MSE performance of Random Forest and XGB is better, with lower errors at 14.4012 and 14.6655, respectively, while CatBoost has the highest MSE at 17.2292. In terms of MAE, Ridge and OLS perform relatively well, with MAEs of 2.9575 and 0.7898, indicating that their average absolute errors during prediction are smaller, outperforming other models, especially CatBoost and MLP, which have larger MAEs.

In terms of the performance of $R^2$, OLS shows an obvious advantage, with a value of 0.9314, much higher than other models, indicating that it has the best fit to the data. On the other hand, Random Forest has the lowest $R^2$ value, close to -0.0131, showing a poor performance.

In summary, the choice of appropriate model depends on specific requirements. Under the condition that the primary code is combined with library API, if researchers focus on training speed, OLS or Ridge can be selected; if researchers pay more attention to prediction accuracy, Random Forest and XGB perform better; in terms of fitting data, OLS is obviously the best choice.

However, it is worth noting that this conclusion is based on data analysts having only the most basic knowledge of Python. This study primarily aims to investigate whether highly efficient DML causal inference models can be created using only the simplest Python code and Library API; and to compare the strengths and weaknesses of each model. Therefore, this result is not intended for researchers of average or higher proficiency.

\section{Discussion}
\subsection{Theoretical Discussion}
The DML theory highlights the significant advantages of Double Machine Learning (DML) in modern causal inference and econometric analysis. Firstly, DML effectively reduces estimation bias by integrating machine learning techniques to control confounding variables, avoiding the biases caused by improper model specification in traditional linear regression models. Secondly, DML has strong adaptability, capable of handling nonlinear relationships and high-dimensional data, making it suitable for complex real-world scenarios, whereas traditional methods rely on strict assumptions and struggle to meet practical needs. Finally, DML performs feature selection and dimensionality reduction through an initial machine learning phase, efficiently addressing the challenges of high-dimensional data and enhancing the accuracy and robustness of causal effect estimation. To sum up, DML outperforms traditional methods in reducing bias, strong adaptability, efficient causal inference, and handling high-dimensional data, becoming a crucial tool in modern causal inference (Chernozhukov et al., 2018).

\subsection{Practical Discussion}
Despite the significant theoretical advantages of DML, it still faces numerous challenges in practical applications. Firstly, model selection is complex, with different models significantly impacting data fitting and generalization capabilities. Incorrect model selection can lead to inaccurate causal estimates. Secondly, DML has high computational complexity, involving training and prediction of multiple machine learning models, which are costly, especially when dealing with large-scale data. Additionally, data quality directly affects the performance of DML; low-quality data can cause model bias, making data preprocessing and cleaning crucial. Feature selection, model interpretability, and generalization ability are also key challenges in DML applications, requiring domain knowledge and data analysis skills to ensure stable performance on new data. Therefore, although DML has advantages in causal inference, it still needs to address challenges such as model selection, computational complexity, data quality, feature selection, interpretability, and generalization ability in practical applications.

\section{Conclusions}
Double machine learning (DML) is effective in high-dimensional causal inference mainly due to its model flexibility, reduced bias, consistency and asymptotic normality, high-dimensional feature selection, as well as model evaluation and robustness. DML reduces bias through a double machine learning framework, flexibly handles complex relationships in high-dimensional data, and uses regularization techniques for feature selection, enhancing the model’s interpretability and predictive power. Additionally, under appropriate conditions, DML ensures the consistency and asymptotic normality of estimators and improves result credibility through robustness tests (Chernozhukov et al., 2018). These characteristics make DML promising for applications in economics, epidemiology, and social sciences (Pearl, Glymour, \& Jewell, 2016).

However, current API still presents certain challenges in practical applications, especially for novice Python users who find it difficult to implement functions with the simplest code and Library API. Many beginners often encounter issues such as unclear documentation, complex API calls, and cumbersome parameter settings when learning and using APIs. These difficulties make them feel confused while writing code, making it hard to smoothly integrate API functionalities into their projects. Therefore, to help novice users better understand and use API technology, more intuitive examples, detailed tutorials, and a friendlier development environment may be needed to lower the learning threshold and improve their programming efficiency.

In addition, during the construction of linear DML models, the mismatch in outcome variable’s dimension is a common issue. This problem often leads to inaccurate results in both training and prediction stages, thereby affecting the final analysis outcomes. Therefore, ensuring that the dimensions of outcome variables match one dimensional array is a critical step. Researchers and practitioners should give this sufficient attention and take appropriate measures to prevent such issues from occurring.

\section{Reference}

\begin{enumerate}
    \item Bach, P., Chernozhukov, V., Kurz, M. S., \& Spindler, M. (2022). \textit{DoubleML – An object-oriented implementation of double machine learning in Python}. Journal of Machine Learning Research, 23(2022), 1–6. Retrieved from \url{http://jmlr.org/papers/v23/21-0862.html}

    \item Chernozhukov, V., Chetverikov, D., Demirer, M., Duflo, E., Hansen, C., Newey, W., \& Robins, J. (2018). \textit{Double/debiased machine learning for treatment and structural parameters}. The Econometrics Journal, 21(1), C1–C68. \url{https://doi.org/10.1111/ectj.12097}

    \item Chernozhukov, V., Chetverikov, D., Demirer, M., Duflo, E., Hansen, C., Newey, W., \& Robins, J. (2024). \textit{Double/Debiased Machine Learning for Treatment and Structural Parameters}. arXiv preprint arXiv:1608.00060v7. Retrieved from \url{https://arxiv.org/abs/1608.00060}

    \item Flores, A. B., \& Chernozhukov, V. (2018). \textit{DoubleML for Python (Version 0.0.1)} [Computer software]. GitHub. Retrieved from \url{https://github.com/DoubleML/doubleml-for-py}

    \item Fuhr, J., Berens, P., \& Papies, D. (2024). \textit{Estimating causal effects with double machine learning – A method evaluation}. School of Business and Economics, University of Tübingen.\\ \url{https://doi.org/10.48550/arXiv.2403.14385}

    \item Pearl, J., Glymour, M., \& Jewell, N. P. (2016). \textit{Causal inference in statistics: A primer}. Wiley.
\end{enumerate}

\section{Appendix}

\subsection{The code for this study}

\subsubsection{Random Forest}
\begin{lstlisting}[language=Python, caption=, label=!]
import numpy as np
import time
from sklearn.metrics import mean_squared_error, mean_absolute_error, r2_score
from sklearn.model_selection import train_test_split
from sklearn.ensemble import RandomForestRegressor
from econml.dml import LinearDML

np.random.seed(42)
X = np.random.randn(1000, 5)
y = X @ np.array([1.5, -2, 0.5, 0, 3]) + np.random.randn(1000)
X_train, X_test, y_train, y_test = train_test_split(X, y, test_size=0.2, random_state=42)

T_train = X_train[:, :3]
X_train_dml = X_train[:, 3:]
T_test = X_test[:, :3]
X_test_dml = X_test[:, 3:]

dml_model = LinearDML(
    model_y=RandomForestRegressor(),
    model_t=RandomForestRegressor(),
    discrete_treatment=False,
    linear_first_stages=False
)

def evaluate_model_dml(dml_model, X_train_dml, y_train, X_test_dml, y_test, T_train, T_test):
    start_time = time.time()
    dml_model.fit(y_train, T_train, X=X_train_dml)
    train_time = time.time() - start_time
    y_pred = dml_model.effect(X_test_dml)
    mse = mean_squared_error(y_test, y_pred)
    mae = mean_absolute_error(y_test, y_pred)
    r2 = r2_score(y_test, y_pred)
    print(f'DML Model Results:')
    print(f'Mean Squared Error (MSE): {mse:.4f}')
    print(f'Mean Absolute Error (MAE): {mae:.4f}')
    print(f'R!$^2$!: {r2:.4f}')
    print(f'Training Time: {train_time:.4f} seconds')
    print('-' * 50)

evaluate_model_dml(dml_model, X_train_dml, y_train, X_test_dml, y_test, T_train, T_test)
\end{lstlisting}

\subsubsection{MLP}
\begin{lstlisting}[language=Python, caption=, label=lst:pythoncode]
!pip install pygam
import numpy as np
import time
from sklearn.metrics import mean_squared_error, mean_absolute_error, r2_score
from sklearn.model_selection import train_test_split
from sklearn.ensemble import RandomForestRegressor
from econml.dml import LinearDML

n = 5000
X = np.random.randn(n, 5)
y = X @ np.array([1.5, -2, 0.5, 0, 3]) + np.random.randn(n)
X_train, X_test, y_train, y_test = train_test_split(X, y, test_size=0.2, random_state=42)

T_train = X_train[:, :3]
X_train_dml = X_train[:, 3:]
T_test = X_test[:, :3]
X_test_dml = X_test[:, 3:]

import statsmodels.api as sm
from sklearn.neural_network import MLPRegressor

dml_model = LinearDML(
    model_y=MLPRegressor(hidden_layer_sizes=(128, 64), activation='relu', solver='adam', max_iter=500),
    model_t=MLPRegressor(hidden_layer_sizes=(128, 64), activation='relu', solver='adam', max_iter=500),
    discrete_treatment=False,
    linear_first_stages=False
)

evaluate_model_dml(dml_model, X_train_dml, y_train, X_test_dml, y_test, T_train, T_test)
\end{lstlisting}

\subsubsection{XGBoost}
\begin{lstlisting}[language=Python, caption=, label=lst:pythoncode]
import numpy as np
import time
from sklearn.metrics import mean_squared_error, mean_absolute_error, r2_score
from sklearn.model_selection import train_test_split
from sklearn.ensemble import RandomForestRegressor
from econml.dml import LinearDML

n = 5000
X = np.random.randn(n, 5)
y = X @ np.array([1.5, -2, 0.5, 0, 3]) + np.random.randn(n)
X_train, X_test, y_train, y_test = train_test_split(X, y, test_size=0.2, random_state=42)

T_train = X_train[:, :3]
X_train_dml = X_train[:, 3:]
T_test = X_test[:, :3]
X_test_dml = X_test[:, 3:]

from xgboost import XGBRegressor
from econml.dml import LinearDML

dml_model = LinearDML(
    model_y=XGBRegressor(n_estimators=200, learning_rate=0.05, max_depth=6),
    model_t=XGBRegressor(n_estimators=200, learning_rate=0.05, max_depth=6),
    discrete_treatment=False,
    linear_first_stages=False
)

evaluate_model_dml(dml_model, X_train_dml, y_train, X_test_dml, y_test, T_train, T_test)
\end{lstlisting}

\subsubsection{Catboost}
\begin{lstlisting}[language=Python, escapechar=!]
from sklearn.multioutput import MultiOutputRegressor
from catboost import CatBoostRegressor
from econml.dml import LinearDML

model_t = MultiOutputRegressor(CatBoostRegressor(iterations=200, learning_rate=0.05, depth=6, verbose=0))
dml_model = LinearDML(
    model_y=CatBoostRegressor(iterations=200, learning_rate=0.05, depth=6, verbose=0),
    model_t=model_t,  
    discrete_treatment=False,
    linear_first_stages=False
)

evaluate_model_dml(dml_model, X_train_dml, y_train, X_test_dml, y_test, T_train, T_test)
\end{lstlisting}

\subsubsection{OLS and SVM}
\begin{lstlisting}[language=Python, escapechar=!]
import numpy as np
import time
from sklearn.metrics import mean_squared_error, mean_absolute_error, r2_score
from sklearn.model_selection import train_test_split, cross_val_score
from sklearn.linear_model import LinearRegression
from sklearn.svm import SVR
from econml.dml import LinearDML
from sklearn.ensemble import RandomForestRegressor

np.random.seed(42)
X = np.random.randn(1000, 5)
y = X @ np.array([1.5, -2, 0.5, 0, 3]) + np.random.randn(1000)
X_train, X_test, y_train, y_test = train_test_split(X, y, test_size=0.2, random_state=42)

def evaluate_model(model, X_train, X_test, y_train, y_test, model_name):
    start_time = time.time()
    model.fit(X_train, y_train)
    train_time = time.time() - start_time
    y_pred = model.predict(X_test)
    mse = mean_squared_error(y_test, y_pred)
    mae = mean_absolute_error(y_test, y_pred)
    r2 = r2_score(y_test, y_pred)
    cv_scores = cross_val_score(model, X_train, y_train, cv=5, scoring='r2')
    cv_mean = cv_scores.mean()
    print(f'{model_name} Results:')
    print(f'Mean Squared Error (MSE): {mse:.4f}')
    print(f'Mean Absolute Error (MAE): {mae:.4f}')
    print(f'R!$^2$!: {r2:.4f}')
    print(f'Cross-Validation R!$^2$!: {cv_mean:.4f}')
    print(f'Training Time: {train_time:.4f} seconds')
    print('-' * 50)

ols_model = LinearRegression()
evaluate_model(ols_model, X_train, X_test, y_train, y_test, 'OLS')

svr_model = SVR(kernel='rbf')
evaluate_model(svr_model, X_train, X_test, y_train, y_test, 'SVM')
\end{lstlisting}

\subsubsection{Lasso}
\begin{lstlisting}[language=Python, escapechar=!]
import numpy as np
import time
from sklearn.metrics import mean_squared_error, mean_absolute_error, r2_score
from sklearn.model_selection import train_test_split
from sklearn.linear_model import LassoCV, MultiTaskLassoCV
from econml.dml import LinearDML

n = 5000
X = np.random.randn(n, 5)
y = X @ np.array([1.5, -2, 0.5, 0, 3]) + np.random.randn(n)
X_train, X_test, y_train, y_test = train_test_split(X, y, test_size=0.2, random_state=42)

T_train = X_train[:, :3]
X_train_dml = X_train[:, 3:]
T_test = X_test[:, :3]
X_test_dml = X_test[:, 3:]

dml_model = LinearDML(
    model_y=LassoCV(cv=5),
    model_t=MultiTaskLassoCV(cv=5),
    discrete_treatment=False,
    linear_first_stages=False
)

evaluate_model_dml(dml_model, X_train_dml, y_train, X_test_dml, y_test, T_train, T_test)
\end{lstlisting}

\subsubsection{Ridge}
\begin{lstlisting}[language=Python, escapechar=!]
import numpy as np
import time
from sklearn.metrics import mean_squared_error, mean_absolute_error, r2_score
from sklearn.model_selection import train_test_split
from sklearn.linear_model import Ridge
from econml.dml import LinearDML

np.random.seed(42)
X = np.random.randn(1000, 5)
y = X @ np.array([1.5, -2, 0.5, 0, 3]) + np.random.randn(1000)
X_train, X_test, y_train, y_test = train_test_split(X, y, test_size=0.2, random_state=42)

T_train = X_train[:, :3]
X_train_dml = X_train[:, 3:]
T_test = X_test[:, :3]
X_test_dml = X_test[:, 3:]

dml_model = LinearDML(
    model_y=Ridge(),
    model_t=Ridge(),
    discrete_treatment=False,
    linear_first_stages=False
)

evaluate_model_dml(dml_model, X_train_dml, y_train, X_test_dml, y_test, T_train, T_test)
\end{lstlisting}

\subsection{Linear DML code based on random forest}
\begin{lstlisting}[language=Python, escapechar=!]
import pandas as pd
from econml.dml import LinearDML
from sklearn.ensemble import RandomForestRegressor
import matplotlib.pyplot as plt
import seaborn as sns
import joblib

data = pd.read_excel('#Your data location')
print(data.head())
print(data.info())

model = LinearDML(
    model_y=RandomForestRegressor(),
    model_t=RandomForestRegressor(),
    discrete_treatment=False,
    linear_first_stages=False
)

model.fit(
    Y=data['y1'],
    T=data[['x1', 'x2', 'x3']],
    X=data[['x4', 'x5']]
)

ate = model.ate(
    X=data[['x4', 'x5']],
    T0=0,
    T1=1
)
print("Average Treatment Effect (ATE):", ate)

cate = model.effect(
    X=data[['x4', 'x5']],
    T0=0,
    T1=1
)
print("Conditional Average Treatment Effect (CATE):", cate)

marginal_effects = model.const_marginal_effect(X=data[['x4', 'x5']])
print("Marginal Effects:", marginal_effects)

ate_effects = model.const_marginal_effect(X=data[['x4', 'x5']])
treatment_vars = ['x1', 'x2', 'x3']
for i, treatment in enumerate(treatment_vars):
    print(f"ATE for {treatment}: {ate_effects[:, i].mean()}")
    print(f"Confidence Interval for {treatment}: "
          f"({ate_effects[:, i].mean() - 1.96 * ate_effects[:, i].std()}, "
          f"{ate_effects[:, i].mean() + 1.96 * ate_effects[:, i].std()})")
\end{lstlisting}

\section{Acknowledgment}

I would like to acknowledge the use of AI tools in the preparation of this research. ChatGPT and DeepSeek were consulted for general suggestions and insights, which were carefully reviewed and refined by the author. Additionally, WPS was utilized for language assistance to improve the clarity and readability of the manuscript. All final decisions regarding the content, analysis, and conclusions were made by the author.

\end{document}